\documentclass{article}
\usepackage{amsmath,amssymb,amsfonts}

\def\op#1{\mathop{{\it\fam0} #1}\limits}

\newcommand{\beq}{\begin{equation}}
\newcommand{\eeq}{\end{equation}}
\newcommand{\ben}{\begin{eqnarray}}
\newcommand{\een}{\end{eqnarray}}
\newcommand{\be}{\begin{eqnarray*}}
\newcommand{\ee}{\end{eqnarray*}}
\newcommand{\bea}{\begin{eqalph}}
\newcommand{\eea}{\end{eqalph}}

\newcommand{\bL}{{\mathbf L}}

\newcommand{\cT}{{\mathcal T}}
\newcommand{\cR}{{\mathcal R}}
\newcommand{\cS}{{\mathcal S}}

\newcommand{\cL}{{\mathcal L}}
\newcommand{\cE}{{\mathcal E}}

\newcommand{\al}{\alpha}
\newcommand{\bt}{\beta}
\newcommand{\dl}{\delta}
\newcommand{\la}{\lambda}

\newcommand{\om}{\omega}

\newcommand{\m}{\mu}
\newcommand{\n}{\nu}
\newcommand{\g}{\gamma}

\newcommand{\ve}{\varepsilon}

\newcommand{\si}{\sigma}
\newcommand{\Si}{\Sigma}

\newcommand{\w}{\wedge}

\newcommand{\wt}{\widetilde}
\newcommand{\wh}{\widehat}
\newcommand{\ol}{\overline}

\newcommand{\dr}{\partial}

\newcommand{\ar}{\op\longrightarrow}

\newcommand{\ot}{\otimes}

\newenvironment{eqalph}{\stepcounter{equation}
\setcounter{equationa}{\value{equation}} \setcounter{equation}{0}

\begin{eqnarray}}{\end{eqnarray}\setcounter{equation}{\value{equationa}}}

\newcounter{equationa}
\newcounter{remark}
\newcounter{example}
\newcounter{theorem}
\newcounter{proposition}
\newcounter{lemma}
\newcounter{corollary}
\newcounter{definition}
\def\theremark{\arabic{remark}}

\def\thedefinition{\arabic{theorem}}

\newenvironment{theorem}{\refstepcounter{theorem} \medskip{\bf
Theorem \thedefinition.}\it}{\medskip }

\newcommand{\mar}[1]{}

\begin{document}

\hbox{}

\begin{center}

{\Large\bf Theory of Classical Higgs Fields. III. Metric-affine
gauge theory}

\bigskip

G. SARDANASHVILY, A. KUROV

\medskip

Department of Theoretical Physics, Moscow State University, Russia

\bigskip

\end{center}

\begin{abstract}
We consider classical gauge theory with spontaneous symmetry
breaking on a principal bundle $P\to X$ whose structure group $G$
is reducible to a closed subgroup $H$, and sections of the
quotient bundle $P/H\to X$ are treated as classical Higgs fields.
Its most comprehensive example is metric-affine gauge theory on
the category of natural bundles where gauge fields are general
linear connections on a manifold $X$, classical Higgs fields are
arbitrary pseudo-Riemannian metrics on $X$, and matter fields are
spinor fields. In particular, this is the case of gauge
gravitation theory.
\end{abstract}

\bigskip
\bigskip
\bigskip

Classical field theory admits a comprehensive mathematical
formulation in the geometric terms of smooth fibre bundles, and
gauge gravitation theory does so \cite{book09,sard08}. In this
framework, classical gauge theory is theory of principal
connections on principal and associated bundles. Following our
previous work \cite{higgs13,higgs14}, we consider classical gauge
theory with spontaneous symmetry breaking on a principal bundle
$P\to X$ whose structure group $G$ is reducible to a closed
subgroup $H$, and sections of the quotient bundle $P/H\to X$ are
treated as classical Higgs fields \cite{book09,sard06a,sard14}. In
this theory, matter fields with an exact symmetry group $H$ are
described by sections of a composite bundle $Y\to P/H\to X$. Their
gauge $G$-invariant Lagrangian necessarily factorizes through a
vertical covariant differential on $Y$ defined by a principal
connection on an $H$-principal bundle $P\to P/H$.

Metric-affine gauge theory provides the most comprehensive example
of classical gauge theory with spontaneous symmetry breaking. In
particular, this is the case of gauge gravitation theory
\cite{heh,iva,obukh}. In a general setting, it is a gauge theory
of general linear connections on natural bundles, examplified by
tangent bundles over a smooth manifold $X$. The associated
principal bundle is a fibre bundle $LX$ of linear frames in
tangent space to $X$. A key point is that its structure group
always is reducible due to the existence of a Riemannian or
pseudo-Riemannian metric on $X$. Thus, a metric-affine gauge
theory necessarily is characterized by spontaneous symmetry
breaking where pseudo-Riemannian metrics are Higgs field. This
fact enables one to describe spinor fields in the framework of
this gauge theory though spinor bundles are not natural. They are
described by sections of the composite bundle (\ref{qqz}). In
particular, the Higgs character of pseudo-Riemannian metrics is
displayed by the fact that the representation (\ref{L4'}) of
tangent holonomic coframes on $X$ by $\g$-matrices on spinor
fields in the presence of different pseudo-Riemannian metrics are
nonequivalent.

One naturally requires that gauge gravitation theory incorporates
Einstein's General Relativity and, therefore, it should be based
on the relativity and equivalence principles reformulated in the
fibre bundle terms \cite{iva,sard11}. In these terms, the
relativity principle states that gauge symmetries of classical
gravitation theory are general covariant transformations. Fibre
bundles possessing general covariant transformations constitute
the category of so called natural bundles \cite{book09,kol}.

A fibre bundle $Y\to X$ is called the natural bundle if there
exists a monomorphism of a group of diffeomorphisms of $X$ to a
group of bundle automorphisms of $Y\to X$ over these
diffeomorphisms. Automorphisms $\wt f$ are called general
covariant transformations of $Y$. Accordingly, there exists a
functorial lift of any vector field $\tau$ on $X$ to a vector
field $\ol\tau$ on $Y$ such that $\tau\mapsto\ol\tau$ is a
monomorphism of the Lie algebra $\cT(X)$ of vector field on $X$ to
that $\cT(T)$ of vector fields on $Y$. This functorial lift
$\ol\tau$ is an infinitesimal generator of a local one-parameter
group of local general covariant transformations of $Y$.

From now on, a smooth manifold $X$ is assumed to be Hausdorff,
second-countable and, consequently, paracompact and locally
compact, countable at infinity. It is $n$-dimensional and
oriented. Given a smooth manifold $X$, its tangent and cotangent
bundles $TX$ and $T^*X$ are endowed with bundle coordinates
$(x^\la,\dot x^\la)$ and $(x^\la,\dot x_\la)$ with respect to
holonomic frames $\{\dr_\la\}$ and $\{dx^\la\}$, respectively.

Natural bundles over $X$ are exemplified by tensor bundles
\mar{sp20}\beq
T=(\op\ot^m TX)\ot(\op\ot^k T^*X) \label{sp20}
\eeq
endowed with holonomic bundle coordinates
$(x^\m,x^{\al_1\cdots\al_m}_{\bt_1\cdots\bt_k})$. Given a vector
field $\tau$ on $X$, its functorial lift onto the tensor bundle
(\ref{sp20}) takes a form
\be \wt\tau =
\tau^\m\dr_\m + [\dr_\nu\tau^{\al_1}\dot
x^{\nu\al_2\cdots\al_m}_{\bt_1\cdots\bt_k} + \ldots
-\dr_{\bt_1}\tau^\nu \dot
x^{\al_1\cdots\al_m}_{\nu\bt_2\cdots\bt_k} -\ldots]\dot \dr
_{\al_1\cdots\al_m}^{\bt_1\cdots\bt_k}, \qquad \dot\dr_\la =
\frac{\dr}{\dr\dot x^\la}.
\ee

The tensor bundles (\ref{sp20}) possess a structure group
\mar{gl4}\beq
GL_n=GL^+(n,\mathbb R). \label{gl4}
\eeq
The associated principal bundle is a frame bundle $LX$ of linear
frames in the tangent spaces to $X$. Given a holonomic bundle
atlas of the tangent bundle $TX$, every element $\{H_a\}$ of a
frame bundle $LX$ takes a form $H_a=H^\m_a\dr_\m$, where $H^\m_a$
is a matrix of the natural representation of a group $GL_4$ in
$\mathbb R^4$. These matrices constitute bundle coordinates
\be
(x^\la, H^\m_a), \qquad H'^\m_a=\frac{\dr x'^\m}{\dr
x^\la}H^\la_a,
\ee
on $LX$ associated to its holonomic atlas
\mar{tty}\beq
\Psi_T=\{(U_\iota, z_\iota=\{\dr_\m\})\}, \label{tty}
\eeq
given by local sections $z_\iota=\{\dr_\m\}$. With respect to
these coordinates, the canonical right action of $GL_n$ on $LX$
reads $GL_4\ni g: H^\m_a\to H^\m_bg^b{}_a$.

A frame bundle $LX$ is equipped with a canonical $\mathbb
R^4$-valued one-form
\mar{b3133'}\beq
\theta_{LX} = H^a_\m dx^\m\ot t_a,\label{b3133'}
\eeq
where $\{t_a\}$ is a fixed basis for $\mathbb R^n$ and $H^a_\m$ is
the inverse matrix of $H^\m_a$.

A frame bundle $LX\to X$ is natural. Any diffeomorphism $f$ of $X$
gives rise to a principal automorphism
\mar{025}\beq
\wt f: (x^\la, H^\la_a)\to (f^\la(x),\dr_\m f^\la H^\m_a)
\label{025}
\eeq
of $LX$ which is its general covariant transformation. Any
$LX$-associated bundle
\be
Y=(LX\times V)/GL_n
\ee
with a typical fibre $V$ also is the natural one. It admits a lift
of any diffeomorphism $f$ of its base to an automorphism
$f_Y(Y)=(\wt f(LX)\times V)/GL_n$.

Let us consider gauge theory of principal connections on a frame
bundle $LX$. They yield linear connections on the associated
bundles, and \emph{vice versa}. In particular, a linear connection
on the tangent bundle $TX\to X$ reads
\mar{B}\beq
K= dx^\la\otimes (\dr_\la +K_\la{}^\m{}_\n \dot x^\n \dot\dr_\m).
\label{B}
\eeq
Its curvature takes a form
\mar{1203}\ben
&& R=\frac12R_{\la\m}{}^\al{}_\bt\dot x^\bt dx^\la\w dx^\m\ot\dot\dr_\al,
\label{1203}\\
&& R_{\la\m}{}^\al{}_\bt = \dr_\la K_\m{}^\al{}_\bt - \dr_\m
K_\la{}^\al{}_\bt + K_\la{}^\g{}_\bt K_\m{}^\al{}_\g -
K_\m{}^\g{}_\bt K_\la{}^\al{}_\g. \nonumber
\een
A torsion of the $K$ (\ref{B}) is defined with respect to the
canonical soldering form $dx^\m\ot\dot\dr_\m$ on $TX$. It reads
\mar{191}\beq
T =\frac12 T_\m{}^\n{}_\la  dx^\la\w dx^\m\ot \dot\dr_\n, \qquad
T_\m{}^\n{}_\la  = K_\m{}^\n{}_\la - K_\la{}^\n{}_\m. \label{191}
\eeq
A linear connection is said to be symmetric if its torsion
(\ref{191}) vanishes, i.e., $K_\m{}^\n{}_\la = K_\la{}^\n{}_\m$.

A principal connection on $LX$, any linear connections (\ref{B})
on $TX$ is represented by a section of the quotient bundle
\mar{015}\beq
C=J^1LX/GL_n\to X, \label{015}
\eeq
where $J^1LX$ is the first order jet manifold of sections of
$LX\to X$ \cite{book09,sard08}. It is an affine bundle modelled
over a vector bundle $T^*X\ot_{TX}VTX\to X$, where $VTX$ is the
vertical tangent bundle of the tangent bundle $TX\to X$. Due to
the canonical splitting of the vertical tangent bundle
\mar{mos163}\beq
VTX=TX\op\times_X TX, \label{mos163}
\eeq
the affine bundle $C\to X$ also is modelled over a vector bundle
\mar{opp}\beq
T^*X\op\ot_{TX}TX\to X. \label{opp}
\eeq
With respect to the holonomic atlas $\Psi_T$ (\ref{tty}), the $C$
(\ref{015}) is provided with the bundle coordinates
\be
(x^\la, k_\la{}^\nu{}_\al), \qquad k'_\la{}^\nu{}_\al=
\left[\frac{\dr x'^\nu}{\dr x^\g} \frac{\dr x^\bt}{\dr x'^\al}
k_\m{}^\g{}_\bt + \frac{\dr x^\bt}{\dr x'^\al}\frac{\dr^2
x'^\nu}{\dr x^\m\dr x^\bt} \right] \frac{\dr x^\m}{\dr x'^\la},
\ee
so that, for any section $K$ of $C\to X$, its coordinates
$k_\la{}^\nu{}_\al\circ K=K_\la{}^\nu{}_\al$ are components of the
linear connection $K$ (\ref{B}). Though the fibre bundle $C\to X$
(\ref{015}) is not $LX$-associated, it is a natural bundle which
admits the functorial lift
\be
 \wt\tau_C = \tau^\m\dr_\m
+[\dr_\nu\tau^\al k_\m{}^\nu{}_\bt - \dr_\bt\tau^\nu
k_\m{}^\al{}_\nu - \dr_\m\tau^\nu k_\nu{}^\al{}_\bt +
\dr_{\m\bt}\tau^\al]\frac{\dr}{\dr k_\m{}^\al{}_\bt}
\ee
of any vector field $\tau$ on $X$.

By the well-known theorem \cite{book09,ste}, the structure group
$GL_n$ (\ref{gl4}) of a principal frame bundle $LX$ always is
reducible to its maximal compact subgroup $SO(n)$. Thus,
spontaneous symmetry breaking in gauge theory on natural bundles
necessarily takes place. Global sections of the corresponding
quotient bundle $LX/SO(n)\to X$ are Riemannian metrics on $X$.

However, in gauge gravitation theory, the equivalence principle
reformulated in geometric terms requires that the structure group
$GL_{n=4}$ (\ref{gl4}) of a frame bundle $LX$ is reducible to a
Lorentz group $SO(1,3)$ \cite{iva,sard11}. Then global sections of
the corresponding quotient bundle
\mar{b3203}\beq
\Si_{\mathrm PR}= LX/SO(1,3) \label{b3203}
\eeq
are pseudo-Riemannian metrics of signature $(+,---)$ on a world
manifold $X$. In Einstein's General Relativity, they are
identified with gravitational fields which thus are treated as
classical Higgs fields.

Let us note that, in any field model on fibre bundles over a
manifold $X$, except the topological ones, a base manifold $X$ is
provided with Riemannian or pseudo-Riemannian metrics. In a
general setting, we therefore assume that the structure group
$GL_n$ (\ref{gl4}) of a principal frame bundle $LX$ is reducible
to a subgroup $SO(m,n-m)$, i.e., it $LX$ contains reduced
principal subbundles with a structure group $SO(m,n-m)$. In
accordance with the well-known theorem (\cite{higgs13}, Theorem
1), there is one-to-one correspondence between these reduced
subbundles $L^gX$ and the global sections $G$ of the corresponding
quotient bundle
\mar{vvv}\beq
\Si=LX/SO(m,n-m) \label{vvv}
\eeq
are pseudo-Riemannian metrics on a manifold $X$. For the sake of
convenience, one usually identifies the quotient bundle $\Si$
(\ref{vvv}) with an open subbundle of the tensor bundle
$\Si\subset \op\vee^2 TX$. Therefore, it can be equipped with
bundle coordinates $(x^\la, \si^{\m\nu})$.

Every pseudo-Riemannian metric $g$ defines an associated bundle
atlas
\mar{lat}\beq
\Psi^h=\{(U_\iota,z_\iota^h=\{h_a\})\} \label{lat}
\eeq
of a frame bundle $LX$ such that the corresponding local sections
$z_\iota^h$ of $LX$ take their values into a reduced subbundle
$L^gX$, and the transition functions of $\Psi^h$ (\ref{lat})
between the frames $\{h_a\}$ are $SO(m,n-m)$-valued. The frames
(\ref{lat}):
\mar{b3211a}\beq
\{h_a =h_a^\m(x)\dr_\m\}, \qquad h_a^\m=H_a^\m\circ z_\iota^h,
\qquad x\in U_\iota, \label{b3211a}
\eeq
are called the tetrad frames. Given the bundle atlas $\Psi^h$
(\ref{lat}), the pull-back
\mar{b3211}\beq
h=h^a\ot t_a=z_\iota^{h*}\theta_{LX}=h_\la^a(x) dx^\la\ot t_a
\label{b3211}
\eeq
of the canonical form $\theta_{LX}$ (\ref{b3133'}) by a local
section $z_\iota^h$ is called the (local) tetrad form. It
determines tetrad coframes
\mar{b3211'}\beq
\{h^a =h^a_\m(x)dx^\m\}, \qquad x\in U_\iota, \label{b3211'}
\eeq
in the cotangent bundle $T^*X$. They are the dual of the tetrad
frames (\ref{b3211a}). The coefficients $h_a^\m$ and $h^a_\m$ of
the tetrad frames (\ref{b3211a}) and coframes (\ref{b3211'}) are
called the tetrad functions. They provide transition functions
between the holonomic atlas $\Psi_T$ (\ref{tty}) and the atlas
$\Psi^h$ (\ref{lat}) of a frame bundle $LX$. Relative to the atlas
(\ref{lat}), a pseudo-Riemannian metric $g$ takes the well-known
form
\mar{mos175}\beq
g=\eta(h\ot h)=\eta_{ab}h^a\ot h^b, \qquad
g_{\m\nu}=h_\m^ah_\nu^b\eta_{ab}, \label{mos175}
\eeq
where $\eta$ is a pseudo-Euclidean metric of signature $(m,n-m)$
in $\mathbb R^n$ written with respect to its fixed basis
$\{t_a\}$.

Since the fibre bundle $C\to X$ is modelled over the vector bundle
(\ref{opp}), given a pseudo-Riemannian metric $g$, any connection
$K$ (\ref{B}) admits a splitting
\mar{mos191}\beq
g_{\n\bt}K_\m{}^\bt{}_\al=K_{\m\n\al}=\{_{\m\n\al}\} +S_{\m\n\al}
+\frac12 C_{\m\n\al} \label{mos191}
\eeq
in the Christoffel symbols
\mar{b1.400}\beq
\{_{\m\n\al}\}= -\frac12(\dr_\m g_{\nu\al} + \dr_\al
g_{\nu\m}-\dr_\nu g_{\m\al}), \label{b1.400}
\eeq
the non-metricity tensor
\mar{mos193}\beq
C_{\m\n\al}=C_{\m\al\n}=D^K_\m g_{\n\al}=\dr_\m g_{\n\al}
+K_{\m\n\al} + K_{\m\al\n} \label{mos193}
\eeq
and the contorsion
\mar{mos202}\beq
S_{\m\n\al}=-S_{\m\al\n}=\frac12(T_{\n\m\al} +T_{\n\al\m} +
T_{\m\n\al}+ C_{\al\n\m} -C_{\n\al\m}), \label{mos202}
\eeq
where $T_{\m\nu\al}=-T_{\al\nu\m}$ are coefficients of the torsion
form (\ref{191}) of $K$.

A connection $K$ is called a metric connection for a
pseudo-Riemannian metric $g$ if $g$ is its integral section, i.e.,
the metricity condition $D^K_\m g_{\n\al}=0$. A metric connection
reads
\mar{mos204}\beq
K^g{}_{\m\n\al}=\{_{\m\n\al}\} + \frac12(T_{\n\m\al} +T_{\n\al\m}
+ T_{\m\n\al}). \label{mos204}
\eeq
By virtue of the well-known theorem \cite{kob}, this connection is
reduced to a principal connection on a reduced principal subbundle
$L^gX$. It follows that, given the decomposition (\ref{mos191})
for any pseudo-Riemannian metric $g$, a linear connection $K$
(\ref{B}) on $LX$ contains the component
\mar{Bg}\beq
K^g= dx^\la\otimes (\dr_\la +K^g{}_\la{}^\m{}_\n \dot x^\n
\dot\dr_\m) \label{Bg}
\eeq
which is a principal connection on $L^gX$ (\cite{higgs14}, Theorem
1). With respect to the atlas $\Psi^h$ (\ref{lat}), the connection
$K^g$ (\ref{Bg}) reads
\mar{Bh}\beq
K^g= dx^\la\otimes (\dr_\la + A_\la{}^b{}_a L_b{}^a), \qquad
A_\la{}^b{}_a = -h^b_\m \dr_\la h^\m_a  + K^g{}_\la{}^\m{}_\nu
h^b_\m h^\nu_a,  \label{Bh}
\eeq
where $L_b{}^a$ are generators of a Lie algebra $so(m,n-m)$ in
$\mathbb R^n$.

In the absence of matter fields, dynamic variables of gauge theory
on natural bundles are linear connections and pseudo-Riemannian
metrics on $X$. Therefore, we call it the metric-affine gauge
theory.  It is a field theory on the bundle product
\mar{ggtt}\beq
Y=\Si\op\times_X C \label{ggtt}
\eeq
coordinated by $(x^\la,\si^{\m\nu}, k_\mu{}^\al{}_\bt)$. Its
particular variant of $n=4$, $m=1$ is metric-affine gravitation
theory \cite{heh,iva,obukh}.

Let us restrict our consideration to first order Lagrangian theory
on $Y$ (\ref{ggtt}). In this case, a configuration space of
metric-affine gauge theory is a jet manifold
\mar{kkl}\beq
J^1Y= J^1\Si_{\mathrm PR}\op\times_X J^1C_{\mathrm W}, \label{kkl}
\eeq
coordinated by $(x^\la,\si^{\m\nu}, k_\mu{}^\al{}_\bt,
\si_\la^{\m\nu}, k_{\la\mu}{}^\al{}_\bt)$.

A first order Lagrangian $L_{\mathrm MA}$ of metric-affine
gravitation theory is a defined as a density
\mar{10130}\beq
L_{\mathrm MA}=\cL_{\mathrm MA}(x^\la,\si^{\m\nu},
k_\mu{}^\al{}_\bt, \si_\la^{\m\nu}, k_{\la\mu}{}^\al{}_\bt)\om,
\qquad \om=dx^1\w\cdots\w dx^n, \label{10130}
\eeq
on the configuration space $J^1Y$ (\ref{kkl}) \cite{book09}. Its
Euler--Lagrange operator takes a form
\mar{eeu}\ben
&& \dl L_{\mathrm MA}= (\cE_{\al\bt} d\si^{\al\bt} + \cE^\m{}_\al{}^\bt
dk_\m{}^\al{}_\bt)\w \om. \label{eeu} \\
&& \cE_{\al\bt} =\left(\frac{\dr}{\dr \si^{\al\bt}}
 - d_\la\frac{\dr}{\dr \si^{\al\bt}_\la}\right)\cL_{\mathrm MA},\qquad
\cE^\m{}_\al{}^\bt =\left(\frac{\dr}{\dr k_\mu{}^\al{}_\bt}
 - d_\la\frac{\dr}{\dr k_{\la\mu}{}^\al{}_\bt}\right)\cL_{\mathrm MA},
 \nonumber\\
&& d_\la=\dr_\la + \si^{\al\bt}_\la \frac{\dr}{\dr \si^{\al\bt}} +
k_{\la\mu}{}^\al{}_\bt \frac{\dr}{\dr k_\mu{}^\al{}_\bt} +
\si^{\al\bt}_{\la\nu} \frac{\dr}{\dr \si^{\al\bt}_\nu} +
k_{\la\nu\mu}{}^\al{}_\bt \frac{\dr}{\dr k_{\nu\mu}{}^\al{}_\bt}.
\nonumber
\een
The corresponding Euler--Lagrange equations read
\mar{10141}\beq
\cE_{\al\bt}=0, \qquad \cE^\m{}_\al{}^\bt=0. \label{10141}
\eeq

The fibre bundle (\ref{ggtt}) is a natural bundle admitting the
functorial lift
\mar{gr3}\ben
&& \wt\tau_{\Si C}=\tau^\m\dr_\m +(\si^{\nu\bt}\dr_\nu \tau^\al
+\si^{\al\nu}\dr_\nu \tau^\bt)\frac{\dr}{\dr \si^{\al\bt}} +
\label{gr3}\\
&& \qquad (\dr_\nu \tau^\al k_\m{}^\nu{}_\bt -\dr_\bt \tau^\nu
k_\m{}^\al{}_\nu -\dr_\mu \tau^\nu k_\nu{}^\al{}_\bt
+\dr_{\m\bt}\tau^\al)\frac{\dr}{\dr k_\mu{}^\al{}_\bt} \nonumber
\een
of vector fields $\tau$ on $X$ \cite{book09}. It is an
infinitesimal generator of general covariant transformations.

By analogy with gauge gravitation theory, the Lagrangian
$L_{\mathrm MA}$ (\ref{10130}) of metric-affine gauge theory is
assumed to be invariant under general covariant transformations.
Its Lie derivative along the jet prolongation $J^1\wt\tau_{\Si C}$
of the vector field $\wt\tau_{\Si C}$ for any $\tau$ vanishes,
i.e.,
\mar{10140}\beq
\bL_{J^1\wt\tau_{\Si C}}L_{\mathrm MA}=0. \label{10140}
\eeq

In order to analyze this condition, let us point out that a first
order jet manifold $J^1C$ of the fibre bundle $C$ (\ref{015})
possesses the canonical splitting
\mar{0101}\ben
&&k_{\la\m}{}^\al{}_\bt =
 \frac12(k_{\la\m}{}^\al{}_\bt - k_{\m\la}{}^\al{}_\bt +
k_\la{}^\g{}_\bt k_\m{}^\al{}_\g -k_\m{}^\g{}_\bt
k_\la{}^\al{}_\g) + \label{0101}\\
&& \qquad \frac12(k_{\la\m}{}^\al{}_\bt +
k_{\m\la}{}^\al{}_\bt - k_\la{}^\g{}_\bt k_\m{}^\al{}_\g +
k_\m{}^\g{}_\bt k_\la{}^\al{}_\g)=\frac12(\cR_{\la\m}{}^\al{}_\bt
+\cS_{\la\m}{}^\al{}_\bt), \nonumber
\een
so that, if $K$ is a section of $C_{\mathrm W}\to X$, then
$\cR_{\la\m}{}^\al{}_\bt\circ J^1K=R_{\la\m}{}^\al{}_\bt$ are
components of the curvature (\ref{1203}). Then the following
assertion is analogous to the well-known Utiyama theorem in
Yang--Mills gauge theory of principal connections
\cite{book09,sard11}.

\begin{theorem} \label{httu1} \mar{httu1} If a first order Lagrangian $L_{\mathrm MA}$
on the configuration space (\ref{kkl}) is invariant under general
covariant transformations and it does not depend on the jet
coordinates $\si^{\al\bt}_\la$ (i.e., derivatives of a metric),
this Lagrangian factorizes through the terms
$\cR_{\la\m}{}^\al{}_\bt$ (\ref{0101}).
\end{theorem}

For instance, let us consider a Lagrangian
\mar{10221}\beq
L=\cR\sqrt{\si}\om =\si^{\m\bt}
\cR_{\la\m}{}^\la{}_{\bt}\sqrt{\si}\om, \qquad
\si=\det(\si_{\m\n}). \label{10221}
\eeq
similar to the Hilbert -- Einstein one in metric-affine
gravitation theory.  The corresponding Euler--Lagrange equations
read
\mar{10180,1}\ben
&& \cE_{\al\bt}=\cR_{\al\bt} -\frac12 \si_{\al\bt} \cR=0, \label{10180}\\
&& \cE^\nu{}_\al{}^\bt=-d_\al(\si^{\nu\bt} \sqrt{\si})
+d_\la(\si^{\la\bt}\sqrt{\si})\dl^\nu_\al + \label{10181}\\
&& \qquad (\si^{\nu\g} k_\al{}^\bt{}_\g -\si^{\la\g}\dl^\nu_\al
k_\la{}^\bt{}_\g - \si^{\nu\bt}k_\g{}^\g{}_\al
+\si^{\la\bt}k_\la{}^\nu{}_\al)\sqrt{\si}=0. \nonumber
\een
The equation (\ref{10180}) is an analogy of the Einstein equations
in metric-affine gravitation theory. The equations (\ref{10181})
 are brought into the form
\mar{10209}\ben
&& \sqrt{\si^{-1}}\si_{\nu\ve}\si_{\bt\m}\cE^\nu{}_\al{}^\bt=
 c_{\al\ve\m}-\frac12 \si_{\m\ve}\si^{\la\g}c_{\al\la\g} -
\si_{\al\ve}\si^{\la\bt}c_{\la\bt\m} + \label{10209} \\
&& \qquad \frac12
\si_{\al\ve}\si^{\la\g}c_{\m\la\g} +
 t_{\m\ve\al} + \si_{\m\ve} t_\al{}^\g{}_\g + \si_{\al\ve} t_\g{}^\g{}_\m
=0. \nonumber
\een
where we introduce the torsion
\be
t_{\m\nu\al}=-t_{\al\nu\m}=\si_{\nu\bt}t_\m{}^\bt{}_\al, \qquad
t_\m{}^\nu{}_\la = k_\m{}^\nu{}_\la - k_\la{}^\nu{}_\m,
\ee
and the non-metricity variables
\be
c_{\m\nu\al}=c_{\m\al\nu}=d_\m \si_{\nu\al}
+k_\m{}^\bt{}_\al\si_{\nu\bt} + k_\m{}^\bt{}_\nu\si_{\bt\al}.
\ee
Substituting the contorsion
\be
s_{\m\nu\al}=-s_{\m\al\nu}=\frac12(t_{\nu\m\al} +t_{\nu\al\m} +
t_{\m\nu\al}+ c_{\al\nu\m} -c_{\nu\al\m}),
\ee
into the equation (\ref{10209}), we obtain the equality
\mar{10183}\ben
&& \frac12 (c_{\al\m\ve} + c_{\m\al\ve}) - \frac12
(\si_{\m\ve}c_\g{}{}^\g{}_\al - \si_{\al\ve}c_\g{}{}^\g{}_\m) + \label{10183}\\
&& \qquad s_{\m\ve\al} - s_{\al\ve\m}+ \si_{\m\ve}(s_\al{}^\g{}_\g - s_\g{}^\g{}_\al)
- \si_{\al\ve}(s_\m{}^\g{}_\g - s_\g{}^\g{}_\m)=0. \nonumber
\een
Its symmetrization with respect to the indices $\al$, $\m$ leads
to the condition $c_{\al\m\ve} + c_{\m\al\ve}=0$, which together
with the equality $c_{\m\nu\al}=c_{\m\al\nu}$ result in that the
non-metricity vanishes:
\mar{789}\beq
c_{\m\nu\al}=_\m \si_{\nu\al} +k_\m{}^\bt{}_\al\si_{\nu\bt} +
k_\m{}^\bt{}_\nu\si_{\bt\al}=0. \label{789}
\eeq
Then the
skew-symmetric part of the equality (\ref{10183}) with respect to
the indices $\al$, $\m$ takes a form
\be
s_{\m\ve\al} - s_{\al\ve\m}+ \si_{\m\ve}(s_\al{}^\g{}_\g -
s_\g{}^\g{}_\al) - \si_{\al\ve}(s_\m{}^\g{}_\g -
s_\g{}^\g{}_\m)=0.
\ee
Its pairing with $\si^{\ve\al}$ leads to the equalities
\be
(s_\al{}^\g{}_\g - s_\g{}^\g{}_\al)=0, \qquad s_{\m\ve\al} -
s_{\al\ve\m}=0,
\ee
and to that both a contorsion $s_{\m\ve\al}$ and a torsion
$t_{\m\nu\al}$ equal zero. Then we obtain from the equality
(\ref{789}) that
\be
k_{\m\n\al}= -\frac12(d_\m \si_{\nu\al} + d_\al \si_{\nu\m}-d_\nu
\si_{\m\al}).
\ee
Substitution of these expression into the equation (\ref{10180})
results in a second order differential equation for the metric
variables $\si$.

Matter fields in metric-affine gauge theory are spinor fields,
e.g., Dirac's fermion fields in gravitation theory, we restrict or
consideration to gravitation theory where they are spinor fields
\cite{book09,sard11}. They do not admit general covariant
transformations, and we therefore follow the procedure in our
previous works \cite{higgs13,higgs14} in order to describe them.

Spinors are conventionally described in the framework of formalism
of Clifford algebras \cite{law}.

Let $V$ be an $n$-dimensional real vector space provided with a
pseudo-Euclidean metric $\eta$. Let us consider a tensor algebra
\be
\otimes V= \mathbb{R} \oplus V\oplus \op\otimes^2V\oplus\cdots
\oplus \op\otimes^k V\oplus \cdots
\ee
of $V$ and its two-sided ideal $I_\eta$ generated by the elements
\be
v\otimes v'+v'\otimes v - 2\eta(v,v')e, \qquad  v,v'\in V,
\ee
where $e$ denotes the unit element of $\otimes V$. The quotient
$\otimes V/I_\eta$ is a real non-commutative ring.

There is the canonical monomorphism of a real vector space $V$ to
$\otimes V/I_\eta$. It is a generating subspace of a real ring
$\otimes V/I_\eta$. A real ring $\otimes V/I_\eta$ together with a
fixed generating subspace $V$ is called the Clifford algebra
$\mathcal{C\ell }(V,\eta)$ modelled over a pseudo-Euclidean space
$(V,\eta)$.

Given Clifford algebras $\mathcal{C\ell }(V,\eta)$ and
$\mathcal{C\ell }(V',\eta')$, by their isomorphism $\mathcal{C\ell
}(V,\eta) \to \mathcal{C\ell }(V',\eta')$ is meant an isomorphism
of them as real rings which also is an isomorphism of their
generating pseudo-Euclidean spaces
\be
\mathcal{C\ell }(V,\eta)\supset (V,\eta)\to
(V',\eta')\subset\mathcal{C\ell }(V',\eta').
\ee
One can show that two Clifford algebras $\mathcal{C\ell }(V,\eta)$
and $\mathcal{C\ell }(V',\eta')$ are isomorphic iff they are
modelled over pseudo-Euclidean spaces $(V,\eta)$ and $(V',\eta')$
of the same signature \cite{law}.

Let a pseudo-Euclidean metric $\eta$ be of signature
$(m;n-m)=(1,...,1;-1,...,-1)$. Let $\{v^1,...,v^n\}$ be a basis
for $V$ such that $\eta$ takes a diagonal form
\be
\eta^{ab}=\eta(v^a,v^b)=\pm \delta^{ab}.
\ee
Then a real ring $\mathcal{C\ell }(V,\eta)$ is generated by the
elements $v^1,...,v^n$ which obey the relations
\be
v^a v^b+v^b v^a=2\eta^{ab}e.
\ee
We agree to call $\{v^1,...,v^n\}$ the basis for a Clifford
algebra $\mathcal{C\ell }(V,\eta)$. Given this basis, let us
denote $\mathcal{C\ell }(V,\eta)=\mathcal{C\ell }(m,n-m)$.

It may happen that a real ring $\mathcal{C\ell }(V,\eta)$ admits a
generating pseudo-Euclidean subspace $(V',\eta')$ of signature
different from that of $(V,\eta)$. In this case, $\mathcal{C\ell
}(V,\eta)$ possesses the structure of a Clifford algebra
$\mathcal{C\ell }(V',\eta')$ which is not isomorphic to a Clifford
algebra $\mathcal A=\mathcal{C\ell }(V,\eta)$. For instance, there
are the following real ring isomorphisms \cite{law}:
\be
\mathcal{C\ell}(m,n-m)\simeq\mathcal{C\ell }(n-m+1,m-1),\quad
\mathcal{C\ell}(m,n-m)\simeq\mathcal{C\ell }(m-4,n-m+4), \quad
n,m\geq 4.
\ee

Let $\mathcal{C\ell }(V,\eta)$ be a Clifford algebra modelled over
a pseudo-Euclidean space $(V,\eta)$, and let Aut$[\mathcal{C\ell
}(V,\eta)]$ denote the group of automorphisms of a real ring
$\mathcal{C\ell }(V,\eta)$. Restricted to $V\subset \mathcal{C\ell
}(V,\eta)$, an automorphism of a real ring $\mathcal{C\ell
}(V,\eta)$ need not be an automorphism of a pseudo-Euclidean space
$(V,\eta)$ and, therefore, it fails to be an automorphism of
$\mathcal{C\ell }(V,\eta)$ as a Clifford algebra in general. An
automorphism $g\in \mathrm{Aut}[\mathcal{C\ell }(V,\eta)]$ is an
automorphism of a Clifford algebra $\mathcal{C\ell }(V,\eta)]$ iff
it is an automorphism of a pseudo-Euclidean space $V$. Conversely,
let $O(V,\eta)$ be a group of automorphisms
\mar{45}\beq
g:V\ni v\mapsto gv\in V, \qquad \eta(gv,gv')=\eta(v,v'), \qquad
g\in O(V,\eta), \label{45}
\eeq
of a pseudo-Euclidean space $(V,\eta)$. Since $\mathcal{C\ell
}(V,\eta)$ is generated by elements of $V$, the action (\ref{45})
of a group $O(V,\eta)$ in $(V,\eta)$ yields automorphisms of a
Clifford algebra $\mathcal{C\ell }(V,\eta)$ so that there exists a
canonical monomorphism
\mar{82}\beq
O(V,\eta)\to \mathrm{Aut}[\mathcal{C\ell }(V,\eta)\,]. \label{82}
\eeq
Herewith, an automorphism of a real ring $\mathcal{C\ell
}(V,\eta)$ is the identity one iff its restriction to $V$ is
$\mathrm{Id}\,V$. Consequently, the following is true.

\begin{theorem} \label{a3} \mar{a3}
A subgroup $O(V,\eta)\subset \mathrm{Aut}[\mathcal{C\ell
}(V,\eta)\,]$ (\ref{82}) exhausts all automorphisms of a Clifford
algebra $\mathcal{C\ell }(V,\eta)$.
\end{theorem}

Invertible elements of a Clifford algebra $\mathcal{C\ell
}(V,\eta)$ constitute a group $\mathcal{GC\ell }(V,\eta)$. Acting
in $\mathcal{C\ell }(V,\eta)$ by left and right multiplications,
this group also acts in a Clifford algebra by the adjoint
representation
\mar{a5}\beq
\mathcal{C\ell }(V,\eta)\supset \mathcal{GC\ell }(V,\eta)\ni g:
a\mapsto gag^{-1}, \qquad a\in \mathcal{C\ell }(V,\eta).
\label{a5}
\eeq
This representation provides a homomorphism
\mar{83}\beq
\zeta:\mathcal{GC\ell }(V,\eta) \to \mathrm{IAut}[\mathcal{C\ell
}(V,\eta)\,](V,\eta)\subset \mathrm{Aut}[\mathcal{C\ell
}(V,\eta)\,] \label{83}
\eeq
of a group $\mathcal{GC\ell }(V,\eta)$ to a subgroup
$\mathrm{IAut}[\mathcal{C\ell }(V,\eta)\,]$ of inner automorphisms
of a real ring $\mathcal{C\ell }(V,\eta)$.

It is readily observed that a group $\mathcal{GC\ell }(V,\eta)$
contains all elements $v\in V\subset \mathcal{C\ell }(V,\eta)$
such that $\eta(v,v)\neq 0$. Then let us consider a subgroup
$\mathrm{Cliff}(V,\eta)\subset \mathcal{GC\ell }(V,\eta)$
generated by all invertible elements of $V\subset \mathcal{C\ell
}(V,\eta)$. It is called the Clifford group.

\begin{theorem} \mar{260} \label{260}
If $n=\dim V$ is even, the homomorphism $\zeta$ (\ref{83}) of a
Clifford group $\mathrm{Cliff}(V,\eta)$ to
$\mathrm{IAut}[\mathcal{C\ell }(V,\eta)]$ is its epimorphism
\mar{103}\beq
\zeta: \mathcal{GC\ell }(V,\eta) \supset \mathrm{Cliff}(V,\eta)\to
O(V,\eta) \subset \mathrm{IAut}[\mathcal{C\ell }(V,\eta)].
\label{103}
\eeq
onto $O(V,\eta)$ \cite{law}.
\end{theorem}

Unless otherwise stated, let us restrict our consideration to
Clifford algebras modelled over an even-dimensional
pseudo-Euclidean space $V$.

The epimorphism (\ref{103}) yields an action of a Clifford group
$\mathrm{Cliff}(V,\eta)$ in a pseudo-Euclidean space $(V,\eta)$ by
the adjoint representation (\ref{a5}). However, this action is not
effective. Therefore, one consider subgroups Pin$(V,\eta)$ and
Spin$(V,\eta)$ of $\mathrm{Cliff}(V,\eta)$. The first one is
generated by elements $v\in V$ such that $\eta(v,v)=\pm 1$. A
group Spin$(V,\eta)$ is defined as an intersection
\be
\mathrm{Spin}(V,\eta)=\mathrm{Pin}(V,\eta)\cap \mathcal{C\ell
}^0(V,\eta)
\ee
of a group Pin$(V,\eta)$ and the even subring $\mathcal{C\ell
}^0(V,\eta)$ of a Clifford algebra $\mathcal{C\ell }(V,\eta)$.  In
particular, generating elements $v\in V$ of Pin$(V,\eta)$ do not
belong to its subgroup Spin$(V,\eta)$.

\begin{theorem} \mar{t1} \label{t1}
The epimorphism (\ref{103}) restricted to the Spin group leads to
short exact sequence of groups
\mar{104}\beq
e\to \mathbb Z_2\longrightarrow
\mathrm{Spin}(V,\eta)\op\longrightarrow^\zeta SO(V,\eta)\to e,
\label{104}
\eeq
where $\mathbb Z_2\to (e,-e)\subset \mathrm{Spin}(V,\eta)$.
\end{theorem}

It should be emphasized that an epimorphism $\zeta$ in (\ref{104})
is not a trivial bundle unless $\eta$ is of signature $(1,1)$. It
is a universal coverings over each component of $O(V,\eta)$.

Let us consider the complexification
\mar{63}\beq
\mathbb C\mathcal{C\ell }(m,n-m)=\mathbb C\op\otimes_{\mathbb
R}\mathcal{C\ell }(m,n-m) \label{63}
\eeq
of a real ring $\mathcal{C\ell }(m,n-m)$. It is readily observed
that all complexifications $\mathbb C\mathcal{C\ell }(m,n-m)$,
$m=0,\ldots, n$, are isomorphic:
\mar{66}\beq
\mathbb C\mathcal{C\ell }(m,n-m)\simeq \mathbb C\mathcal{C\ell
}(m',n-m'), \label{66}
\eeq
both as real and complex rings. Namely, with the bases $\{v^i\}$
and $\{e^i\}$ for $\mathcal{C\ell }(m,n-m)$ and $\mathcal{C\ell
}(n,0)$, their isomorphisms (\ref{66}) are given by associations
\mar{65}\beq
v^{1,\ldots,m}\mapsto e^{1,\ldots,m}, \qquad
v^{m+1,\ldots,n}\mapsto ie^{m+1,\ldots,n}. \label{65}
\eeq

Though the isomorphisms (\ref{65}) are not unique, one can speak
about an abstract complex ring $\mathbb C\mathcal{C\ell }(n)$ so
that, given a real Clifford algebra $\mathcal{C\ell }(m,n-m)$ and
its complexification $\mathbb C\mathcal{C\ell }(m,n-m)$
(\ref{63}), there exist the complex ring isomorphism (\ref{65}) of
$\mathbb C\mathcal{C\ell }(m,n-m)$ to $\mathbb C\mathcal{C\ell
}(n)$. We agree to call $\mathbb C\mathcal{C\ell }(n)$ the complex
Clifford algebra, and to define it as a complex ring
\mar{a25}\beq
\mathbb C\mathcal{C\ell }(n)=\mathbb C\op\otimes_{\mathbb
R}\mathcal{C\ell }(n,0) \label{a25}
\eeq
generated by the basis elements $(\{e^i\}$ for $\mathcal{C\ell
}(n,0)$. The complex ring $\mathbb C\mathcal{C\ell }(n)$
(\ref{a25}) possesses a canonical real subring $\mathcal{C\ell
}(m,n-m)$ with a basis $\{e^1,\ldots, e^m, ie^{m+1}, \ldots,
ie^n\}$.

Let $\mathbb C\mathcal{C\ell}(n)$ be a complex Clifford algebra.
Automorphisms of its real subrings $\mathcal{C\ell}(m,n-m)$ yields
automorphisms of $\mathbb C\mathcal{C\ell}(n)$, but do not exhaust
all automorphisms of $\mathbb C\mathcal{C\ell}(n)$.

By a representation of a complex Clifford algebra $\mathbb
C\mathcal{C\ell }(n)$ is meant its morphism $\g$ to a complex
algebra of linear endomorphisms of a finite-dimensional complex
vector space.

\begin{theorem} \label{a11} \mar{a11}
If $n$ is even, an irreducible representation of a complex
Clifford algebra $\mathbb C\mathcal{C\ell }(n)$ is unique up to an
equivalence \cite{law}. If $n$ is odd there exist two
non-equivalent irreducible representations of a complex Clifford
algebra $\mathbb C\mathcal{C\ell }(n)$ (see Section 2.5).
\end{theorem}

A complex spinor space $\Xi(n)$ is defined as a carrier space of
an irreducible representation of a complex Clifford algebra
$\mathbb C\mathcal{C\ell }(n)$ is . If $n$ is even, it is unique
up to an equivalence in accordance with Theorem \ref{a11}.
Therefore, it is sufficient to describe a complex spinor space
$\Xi(n)$ as a subspace of a complex Clifford algebra $\mathbb
C\mathcal{C\ell }(n)$ which acts on $\Xi(n)$ by left
multiplications.

Given a complex Clifford algebra $\mathbb C\mathcal{C\ell }(n)$,
let us consider its non-zero minimal left ideal which
$\mathcal{C\ell }(n)$ acts on by left multiplications. It is a
finite-dimensional complex vector space. Therefore, an action of a
complex Clifford algebra $\mathbb C\mathcal{C\ell }(n)$ in a
minimal left ideal by left multiplications defines a linear
representation of $\mathbb C\mathcal{C\ell }(n)$. It obviously is
irreducible. In this case, a minimal left ideal of $\mathbb
C\mathcal{C\ell }(n)$ is a complex spinor space $\Xi(n)$.

Hereafter, we define complex spinor spaces $\Xi(n)$ just as a
minimal left ideals of a complex Clifford algebra $\mathbb
C\mathcal{C\ell }(n)$ which carry out its irreducible
representation. In particular, this definition enables us to
investigate morphisms of complex spinor spaces yielded by
automorphisms of a complex Clifford algebra.

In order to describe complex spinor spaces defined in this way,
one is based the following fact \cite{law}.

\begin{theorem} \mar{a41} \label{a41}
Any complex spinor space $\Xi(n)$ is generated by some Hermitian
idempotent $p\in \Xi(n)$, i.e., $p=p^*$, $p^2=p$, namely,
\be
p=\frac12(e + s), \qquad s^2=e, \qquad s^*=s, \qquad s\neq e.
\ee
\end{theorem}

An irreducible representation of a complex Clifford algebra
$\mathbb C\mathcal{C\ell }(n)$ in a complex spinor space $\Xi(n)$
also implies the representations of both a real Clifford algebra
$\mathcal{C\ell }(m,n-m)$ and the Spin group
$\mathrm{Spin}(m,n-m)$ in $\Xi(n)$ though they need not be
irreducible.

In classical field theory, spinor fields are described by sections
of a spinor bundle on an $n$-dimensional manifold $X$ whose
typical fibre is a spinor space $\Xi(n)$ which carries out a
representation of a complex Clifford algebra $\mathbb
C\mathcal{C\ell }(n)$.

Namely, let $CX\to X$ be a bundle in complex Clifford algebras
$\mathbb C\mathcal{C\ell }(n)$ whose structure group is a group of
automorphisms $\mathrm{Aut}[\mathbb C\mathcal{C\ell }(n)\,]$ of
$\mathbb C\mathcal{C\ell }(n)$. However, $CX$ need not contain a
spinor subbundle because a spinor subspace $\Xi(n)$. being a left
ideal of $\mathbb C\mathcal{C\ell }(n)$, is not stable under
automorphisms of $\mathbb C\mathcal{C\ell }(n)$. A spinor
subbundle $\Xi X\subset CX$ exists if $CX$ also is a fibre bundle
with a structure group $\mathcal{G}G_{\mathrm s}$ of invertible
elements of $\mathbb C\mathcal{C\ell }(n)$ acting on $\mathbb
C\mathcal{C\ell }(n)$ by left multiplications. In this case, a
structure group of $CX$ is reducible to a subgroup $G_{\mathrm s}$
of inner automorphisms of $\mathbb C\mathcal{C\ell }(n)$ generated
by elements of $\mathcal{G}G_{\mathrm s}$ similarly to the
expression (\ref{a5}). Then a representation bundle morphism
\mar{999}\beq
\g: CX\op\times_X CX\supset CX\op\times_X \Xi X\to \Xi X
\label{999}
\eeq
is defined.

A key point is that, in order to construct the Dirac operator on
spinor fields, one need a representation of covectors to $X$ as
elements of a Clifford algebra $\mathbb C\mathcal{C\ell }(n)$
acting in a spinor space $Psi(n)$. A necessary condition of such a
representation is that a structure group $GL_n$ of the cotangent
bundle $T^*X$ over $X$ is reduced to a structure group
$\mathcal{G}G_{\mathrm s}$ of $CX$, i.e., $\mathcal{G}G_{\mathrm
s}$ is some group $SO(m,n-m)$.

Therefore, let a structure group $GL-n$ of a linear frame bundle
be reduced to a subgroup $SO(m,n-m)$, and let $g$ be the
corresponding pseudo-Riemannian metric on $X$ as a global section
of the quotient bundle (\ref{vvv}). In this case, one can think of
the cotangent bundle $T^*X$ over $X$ as being a fibre bundle $MX$
in pseudo-Euclidean spaces with a typical fibre $M$ and a
structure group $SO(m,n-m)$. This fibre bundle is extended to a
fibre bundle in complex Clifford algebras $CX$ whose fibres $C_xX$
are complex Clifford algebras $\mathbb C\mathcal{C\ell }(n)$ which
are complexifications of real Clifford algebras $\mathcal{C\ell
}(M_x,g(x))$ modelled over pseudo-Euclidean tangent spaces
$M_x=T^*_xX$ to $X$. A fibre bundle $CX$ possesses a structure
group $SO(m,n-m)$ of automorphisms of $\mathbb C\mathcal{C\ell
}(n)$, and there is a bundle monomorphism $T^*X\to CX$. In order
to provide the representation (\ref{999}) and in accordance with
the exact sequence of groups (\ref{104}), let us assume that a
structure group $SO(m,n-m)$ of $CX$ lifts to a structure Spin
group $\mathrm{Spin}(m,n-m)$. By virtue of the well-known theorem
\cite{book09,greub}, this can occur if the second Stiefel --
Whitney class of $X$ is trivial. Then we get the representation
morphism
\mar{998}\beq
CX\op\times_X CX\supset T^*X\op\times_X \Xi X\to \Xi X.
\label{998}
\eeq

Thus, we come to the following notion \cite{book09,law}. A spinor
structure on a manifold $X$ is defined as a pair $(P^g, z_s)$ of a
principal bundle $P^g\to X$ with a structure Spin group
$G_{\mathrm{s}}=\mathrm{Spin}(m,n-m)$ and its bundle morphism
$z_s: P^g \to LX$ to a frame bundle $LX$. This morphism factorizes
\mar{g10}\beq
z_s: P^g \to L^gX\subset LX \label{g10}
\eeq
through some reduced principal subbundle $L^gX\subset LX$ with a
structure group $SO(m,n-m)$. Thus, any spinor structure on a
manifold $X$ is associated with a pseudo-Riemannian structure on
$X$. There is the well-known topological obstruction to the
existence of a spinor structure \cite{ger,book09}. To satisfy the
corresponding topological conditions we further assume that a
manifold $X$ is non-compact and parallelizable, that is a linear
frame bundle is trivial.  In this case, all spinor structures
(\ref{g10}) are isomorphic. Therefore, there is one-to-one
correspondence (\ref{g10}) between the pseudo-Riemannian
structures $L^gX$ and the spinor structures $(P^g, z_s)$ which
factorize through the corresponding $L^gX$. We agree to call $P^g$
the spinor principal bundles.

Let us describe the representation morphism (\ref{998}) in an
explicit form.

Due to the factorization (\ref{g10}), every bundle atlas
$\Psi^h=\{z^h_\iota\}$ (\ref{lat}) of $L^gX$ gives rise to an
atlas
\mar{atl0}\beq
\ol\Psi^h=\{\ol z^h_\iota\}, \qquad z^h_\iota =z_h\circ \ol
z^h_\iota, \label{atl0}
\eeq
of a principal $G_{\mathrm s}$-bundle $P^g$.

Let $(P^g, z_h)$ be the spinor structure associated with a
pseudo-Riemannian metric $g$. Let
\mar{y1}\beq
S^g=(P^g\times \Xi(n))/G_{\mathrm s}\to X \label{y1}
\eeq
be a $P^g$-associated spinor bundle whose typical fibre $\Xi(n)$
carriers a representation of a Spin group group $G_{\mathrm s}$.
Given the atlas (\ref{atl0}), let $S^g$ (\ref{y1}) be provided
with the associated bundle coordinates $(x^\m, y^A)$.

Let us consider an $L^gX$-associated bundle in pseudo-Riemannian
spaces
\mar{int2}\beq
M^gX=(L^gX\times M)/SO(m,n-m)=(P^g\times M)/G_{\mathrm s}.
\label{int2}
\eeq
It is isomorphic to the cotangent bundle
\be
T^*X=(L^gX\times M)/SO(m,n-m).
\ee
Then, using the morphism (\ref{998}), one can define a
representation
\mar{L4}\beq
\g_g: T^*X\op\times_X S^g \to S^g \label{L4}
\eeq
of covectors to $X$ by Dirac $\g$-matrices on elements of a spinor
bundle $S^h$. Relative to a bundle atlas $\{z^h_\iota\}$ of $LX$
and the corresponding atlas $\{\ol z_\iota\}$ (\ref{atl0}) of a
spinor principal bundle $P^g$, the representation (\ref{L4}) reads
\be
y^A(\g_g(h^a(x))=\g(t^a)^A{}_By^B,
\ee
where $y^A$ are associated bundle coordinates on $S^g$, and $h^a$
are tetrad coframes. For brevity, we write
\mar{L4'}\beq
\wh h^a=\g_g(h^a)=\g(t^a)=\g^a,\qquad \wh
dx^\la=\g_g(dx^\la)=h^\la_a(x)\g^a. \label{L4'}
\eeq

Given the representation (\ref{L4'}), one can introduce a Dirac
operator on $S^g$ with respect to a principal connection on $P^g$.
Then one can think of sections of $S^g$ (\ref{y1}) as describing
spinor fields in the presence of a pseudo-Riemannian metric $g$.

Note that there is one-to-one correspondence between the principal
connections on $P^g$ and those on a reduced bundle $L^gX$. It
follows that any linear connection $K$ (\ref{B}), generating the
connection $K_g$ (\ref{Bh}) on $L^gX$, yields the corresponding
spinor connection $K_s$ on $P^g$ and $S^g$ \cite{book09,sard98a}.
This connection takes a form
\mar{ddd}\beq
K_s = K^g= dx^\la\otimes (\dr_\la + A_\la{}^b{}_a I_b{}^a), \qquad
A_\la{}^b{}_a = -h^b_\m \dr_\la h^\m_a  + K^g{}_\la{}^\m{}_\nu
h^b_\m h^\nu_a, \label{ddd}
\eeq
where $I_b{}^a$ are generators of the Lie algebra of a Spin group
$\mathrm{Spin}(m,n-m)$ in a spin space $\Xi(n)$.

This fact enables one to describe spinor fields in the framework
of metric-affine gauge theory with general linear connections.

Spinor fields in the presence of different pseudo-Riemannian
metrics $g$ and $g'$ are described by sections of different spinor
bundles $S^g$ and $S^{g'}$. A problem is that, though reduced
subbundles $L^gX$ and $L^{g'}X$ are isomorphic, the associated
structures of bundles in pseudo-Riemannian spaces $M^gX$ and
$M^{g'}X$ (\ref{int2}) on the cotangent bundle $T^*X$ are
non-equivalent because of non-equivalent actions of a group
$SO(m,n-m)$ on a typical fibre of $T^*X$ seen both as a typical
fibre of $M^gX$ and that of $M^{g'}X$. As a consequence, the
representations $\g_h$ and $\g_{h'}$ (\ref{L4'}) for different
metrics $g$ and $g'$ are non-equivalent \cite{book09,sard98a}.
Indeed, let
\be
 t^*=t_\m dx^\m=t_ah^a=t'_a{h'}^a
\ee
be an element of $T^*X$. Its representations $\g_g$ and $\g_{g'}$
(\ref{L4}) read
\be
\g_g(t^*)=t_a\g^a=t_\m h^\m_a\g^a, \qquad
\g_{g'}(t^*)=t'_a\g^a=t_\m {h'}^\m_a\g^a.
\ee
They are non-equivalent because no isomorphism $\Phi_{\mathrm s}$
of $S^g$ onto $S^{g'}$ can obey the condition
\be
\g_{g'}(t^*)=\Phi_{\mathrm s} \g_g(t^*)\Phi_{\mathrm s}^{-1},
\qquad t^*\in T^*X.
\ee

Since the representations (\ref{L4'}) for different metrics fail
to be equivalent, one meets a problem of describing spinor fields
in the presence of different pseudo-Riemannian metrics.

In order to solve this problem, we follow the procedure in
\cite{higgs14}. Let us consider a universal two-fold covering
$\wt{GL}_4$ of a group $GL_4n$ and a $\wt{GL}_g$-principal bundle
$\wt{LX}\to X$ which is a two-fold covering bundle  of a frame
bundle $LX$ {\cite{book09,law}. Then we have a commutative diagram
\be
\begin{array}{ccc}
 \wt{LX} & \ar^\zeta & LX \\
 \put(0,-10){\vector(0,1){20}} &
& \put(0,-10){\vector(0,1){20}}  \\
P^g & \ar & L^gX
\end{array}
\ee
for any spinor structure (\ref{g10}) \cite{book09,sard98a}. As a
consequence,
\be
\wt{LX}/G_\mathrm{s}=LX/SO(m,n-m)=\Si.
\ee
Since $\wt{LX}\to \Si$ is an $G_\mathrm{s}$-principal bundle, one
can consider an associated spinor bundle $S\to \Si$ whose typical
fibre is $\Xi(n)$. We agree to call it the universal spinor bundle
because, given a pseudo-Riemannian metric $g$, the pull-back
$S^g=g^*S\to X$ of $S$ onto $X$ is a spinor bundle $S^g$ on $X$
which is associated with an $G_\mathrm{s}$-principal bundle $P^g$.
A spinor bundle $S$ is endowed with bundle coordinates $(x^\la,
\si^\m_a,y^A)$, where $(x^\la, \si^\m_a)$ are bundle coordinates
on $\Si$ and $y^A$ are coordinates on a spinor space $\Xi(n)$. A
spinor bundle $S\to\Si$ is a subbundle of a bundle in Clifford
algebras which is generated by a bundle of pseudo-Riemannian
spaces associated with a $SO(m,n-m)$-principal bundle $LX\to\Si$.
As a consequence, there is a representation
\mar{L7}\beq
\g_\Si: T^*X\op\times_\Si S \to S, \qquad \g_\Si (dx^\la)
=\si^\la_a\g^a, \label{L7}
\eeq
whose restriction to a subbundle $S^g\subset S$ restarts the
representation (\ref{L4'}).

Sections of a composite bundle
\mar{qqz}\beq
S\to \Si\to X \label{qqz}
\eeq
describe spinor fields in the presence of different
pseudo-Riemannian metrics as follows.

By virtue of (\cite{higgs14}, Theorem 6), any linear connection
$K$ on $X$ (\ref{B}) yields a connection
\mar{b3266}\beq
A_\Si = dx^\la\ot(\dr_\la   + K^g{}_\la{}^\m{}_\nu \si^b_\m
\si^\nu_a I_b{}^a) + d\si^b_\m \ot( \dr_\la -si^\m_a I_b{}^a)
 \label{b3266}
\eeq
on the spinor bundle $S\to\Si$ whose restriction to $S^g$ is the
spinor connection $K_\mathrm{s}$ (\ref{ddd}) defined by defined by
$K$. The connection (\ref{b3266}) yields the so called vertical
covariant differential
\be
\wt D= dx^\la\ot[y^A_\la +(\si^b_{\la\m}\si^\m_a -
K^g{}_\la{}^\m{}_\nu \si^b_\m \si^\nu_a)(I_b{}^a)^A{}_By^B],
\ee
on a fibre bundle $S\to X$ (\ref{qqz}). Its restriction to
$J^1S^g\subset J^1S$ recovers the familiar covariant differential
on a spinor bundle $S^g\to X$ relative to the spin connection
(\ref{ddd}).

\end{document}